# Electromagnetic Window into the Dawn of Black Holes

## A White Paper for the 2020 Decadal Survey

**Thematic Areas:**

☐ Planetary Systems
☑ Formation and evolution of compact objects
☐ Stars and Stellar Evolution
☑ Galaxy Evolution
☐ Star and Planet formation
☑ Cosmology and Fundamental Physics
☐ Resolved stellar populations and their environments
☑ Multi-Messenger Astronomy and Astrophysics

**Authors:** Zoltán Haiman (*Columbia University*), William N. Brandt (*Pennsylvania State University*), Alexey Vikhlinin (*Harvard-Smithsonian Center for Astrophysics*), Jillian Bellovary (*Queensborough Community College/AMNH*), Elena Gallo (*University of Michigan*), Jenny Greene (*Princeton University*), Kohei Inayoshi (*Peking University*), Joseph Lazio (*JPL*), Bret Lehmer (*University of Arkansas*), Bin Luo (*Nanjing University*), Piero Madau (*UC Santa Cruz*), Priya Natarajan (*Yale University*), Feryal Özel (*University of Arizona*), Fabio Pacucci (*Kapteyn Astronomical Institute*), Alberto Sesana (*University of Birmingham*), Daniel Stern (*JPL*), Christian Vignali (*University of Bologna*), Eli Visbal (*Flatiron Institue*), Fabio Vito (*Pontifícia Universidad Católica de Chile*), Marta Volonteri (*Institut d'Astrophysique de Paris*), Joan Wrobel (*NRAO*)

**Endorsers:** Emanuele Berti (*Johns Hopkins University*), Volker Bromm (*University of Texas*), Greg Bryan (*Columbia University*), Nico Cappelluti (*University of Miami*), Anastasia Fialkov (*University of Sussex*), Martin Haehnelt (*Cambridge University*), John Regan (*Dublin City University*), Angelo Ricarte (*Yale University*), John Wise (*Georgia Institute of Technology*), Jemma Wolcott-Green (*Columbia University*)

**Contact:** Zoltán Haiman
*Address:* Department of Astronomy
Mail Code 5246, Columbia University
New York, NY 10027, USA

*Email:* zoltan@astro.columbia.edu

*Phone:* (212) 854-6822

*Executive Summary.*  Massive black holes, $M_{BH} = 10^6 - 10^{10}\ M_\odot$, are known to reside in local galactic nuclei. Observations over the past 20 years have revealed that these black holes are common by the time the Universe is several Gyr old. Many of them were in place very quickly, within the first 1 Gyr after the Big Bang. Their quick assembly has been attributed to mechanisms such as the rapid collapse of gas into the nuclei of early protogalaxies, accretion and mergers of stellar-mass black holes accompanying hierarchical structure formation at early times, and the runaway collapse of early, ultra-dense stellar clusters. The origin of the early massive black holes remains one of the most intriguing and longest-standing unsolved puzzles in astrophysics.

**In this white paper, we discuss strategies for discerning between black hole seeding models using *electromagnetic* observations.** We argue that the most direct answers will be obtained through detection of black holes with masses of $\lesssim 10^5\ M_\odot$ at the redshifts $z \gtrsim 10$ where we expect them to form. Reaching out to these redshifts and down to these masses is crucial, because black holes are expected to "lose" the memory of their initial assembly by the time they grow well above $\sim 10^5\ M_\odot$ and are incorporated into higher-mass galaxies.

The best way to detect a population of $10^{4-5}\ M_\odot$ black holes at high redshifts is by a sensitive X-ray survey. Critical constraining power is augmented by establishing the properties and the environments of their host galaxies in deep optical & IR imaging surveys. Required optical & infrared (OIR) data can be obtained with the already planned *JWST* and *WFIRST* missions. The required X-ray flux limits (down to $\sim 10^{-19}$ erg s$^{-1}$ cm$^{-2}$) are accessible only with a next-generation X-ray observatory which has both high angular resolution ($\lesssim 1''$) and high throughput.

A combination of deep X-ray and OIR surveys we describe below will be capable of probing several generic "markers" of the black hole seed scenarios. Thus it will resolve the long-stanging puzzle of their origin. These electromagnetic observations are also extremely synergistic with the information provided by *LISA* detections of high-$z$ black hole mergers [1].

*Demographics of Supermassive Black Holes, Theories of Their Origin, Main Questions.*  The local population of inactive supermassive black holes (SMBHs) is now well studied down to $\sim 10^6\ M_\odot$ mass scales [2]. SMBHs with $M = 10^6 - 10^{10}\ M_\odot$ are commonly found in the nuclei of nearby major galaxies. About 1% are active quasars. It is believed that ubiquitous SMBHs in local galaxies are the remnants of high-$z$ quasars, which assembled most of their mass in brief ($10^{7-8}$ yrs) but bright episodes of accretion at $z \sim 2 - 3$ [3–5]. However, the physical mechanism through which SMBHs acquired the first $\sim 10^6\ M_\odot$ of their mass remains a mystery [6, 7] and is commonly referred to as the supermassive black hole seed problem. This is a major puzzle because at least some SMBHs with $M = 10^8 - 10^9\ M_\odot$ are in place as early as redshift $z = 7.5$ [8–13], $\lesssim 700$ Myr after the Big Bang. Several distinct physical mechanisms to form high-redshift $\sim 10^5 M_\odot$ SMBHs, corresponding to just below the low-end of the known observed range, are under consideration [6]:

1. Sustained Eddington-rate accretion onto an initially stellar-mass BH. (*"Light seeds"*).
2. Direct collapse of a gas cloud into a BH seed without fragmentation or star formation.
3. Rapid gas collapse into a BH seed via an intermediary stage of a supermassive star.
4. Rapid gas collapse onto a pre-existing stellar-mass BH at super- or hyper-Eddington rates.
5. Runaway collapse and merger of an ultra-dense stellar cluster.

Options (2)–(5) can all produce a $10^{4-5}\ M_\odot$ seed BH within $< 1$ Myr and are often collectively referred to as *"Heavy seeds"*. Note also that there may be a continuum between the light and heavy seed scenarios, with initially stellar-mass BHs growing at modestly super-Eddington rates [14]. **Understanding which of these mechanisms are responsible for the first SMBHs at $z > 6$ is one of the most important unsolved problems in astrophysics that will be addressed with next-generation facilities** [7, 15–21].

Option (1) is the simplest possibility, which is attractive because we do expect an abundance of 10–100 $M_\odot$ BHs formed as remnants of massive and short-lived PopIII stars as early as $z \sim 30$–$40$ [22]. However, this scenario is challenging [23] on several grounds. It requires sustained accretion near the Eddington rate from $z \sim 30$ through $z \sim 6$. PopIII stars are born in small "minihalos", in which the gas fuel for a future BH is easily unbound by feedback [24–27]. Gravitational radiation recoil during BH mergers can easily eject them from low-mass host halos [28] and thus stunt further growth. Because of challenges with the light seeds scenario, the heavy seeds channels are also attractive. Rapid formation of a massive BH via any of these mechanisms would be among the most spectacular events in the history of the Universe. Therefore, **an equally interesting question is to determine whether *any* of the high-$z$ SMBHs form via a heavy-seed channel.**

All heavy seed scenarios share a common feature: they take place in "atomic cooling halos" (ACHs) at $z \gtrsim 10 - 12$, in which gas cools efficiently via line emission (e.g., Lyman $\alpha$) of atomic hydrogen. Such halos exist in a narrow range of virial temperatures $T_{\rm vir} \approx 8000$K, corresponding to $M_{\rm vir}$ = few $\times 10^7$ $M_\odot$ at $z = 10$. Analytic arguments and numerical models show that rapidly cooling self-gravitating gas collapses into ACHs at rates as high as $\dot{M} \approx 1 M_\odot$ yr$^{-1}$ ($\approx c_s^3/G$ where the sound speed $c_s \approx 10$ km s$^{-1}$ for $T = 8000$ K). This inflow can feed formation of a massive BH seed, but the necessary condition is that the gas is prevented from fragmenting into normal stars [29–31], which is challenging. This can be achieved in a small subset of ACHs by exposing them to extremely intense Lyman-Werner radiation [32–35] from a bright neighbor [36–38], and/or by intense heating from an unusually violent merger history [39–41], aided by the streaming motions between gas and dark matter [42–44]. When these conditions are met, the ACHs can quickly (within a few Myr) produce BHs with masses as high as $10^{4-5}$ $M_\odot$.

Note that *none* of the heavy seed formation channels are expected to promptly yield BHs with masses well above the $10^{4-5}$ $M_\odot$ range. For direct collapse, the total gas supply in the $M_{\rm tot}$ = few $\times 10^7$ $M_\odot$ ACH is limited [45]. The collapse of a supermassive star into a BH sets in via general relativistic instability at $\approx 10^5 M_\odot$ [46–49]. Hyper-Eddington accretion onto a lower-mass BH ceases once a similar mass is reached [50–52]. Finally, an ultra-dense nuclear stellar cluster in the core of the ACH cannot contain more than $10^5$ $M_\odot$ of stellar material [53, 54].

*Observational Diagnostics.*   Observational probes of the light vs. heavy seeds scenarios include:
- *High-$z$ luminosity functions*, expected to show a sharp drop at high $L$ (or $M_{\rm BH}$) in the light-seed case [55] and different shapes of the faint end for different formation channels [56, 57].
- *Characteristic spectral signatures for on-going heavy seed formation events* in the dense cores of ACHs [58, 59], due to unusually large obscuring columns expected in these objects.
- *Event rates, mass and spin distributions of mergers* detectable by *LISA* [55, 60–65].
- *Properties of BH host galaxies*. As discussed above, a shared feature of all heavy-seed scenarios is that the initial SMBH with $M = 10^{4-5} M_\odot$ is born in a few$\times 10^7 M_\odot$ halo whose total gas content is few$\times 10^6 M_\odot$. In the light seed models, similar $10^{4-5} M_\odot$ SMBHs are assembled by numerous mergers and slower accretion, and thus are located inside much more massive hosts by $z = 10$ [66].
- *Cross-correlation of the redshifted 21cm background with the residual X-ray background* [67].
- *Fossil evidence* from the local population of intermediate-mass black holes [68–70].

There are large theoretical uncertainties and degeneracies for each of these probes. Particularly challenging can be predictions for "mixed" models, in which early BHs are formed through both heavy and light seed channels. Therefore, the best route is through a combination of techniques and relying on the most generic markers of different scenarios.

Catching the on-going formation of a heavy seed or its subsequent early phases, while the seed stays in its original host, would yield an exceptionally clean test (e.g. [71–73], see a full exposition in [74]). But even if we typically miss this phase, sensitive, next-generations surveys can still probe the expected differences at both the faint and bright ends of the luminosity function [56, 57, 65, 66, 75–77] and very distinct differences in the abundances and the host galaxy properties. The merger histories of heavy seeds indicate that their host galaxies remain strong outliers for ≳ 100 Myr after the BH birth, with BH-to-stellar mass ratio of order unity. This contrasts sharply with the light seed scenario, in which $M_{BH}/M_\star$ is of order $10^{-3}$, consistent with the local relation [73].

A key requirement for any observational test of seed models is to **detect BHs with mass ~ $10^5 M_\odot$ or below at $z > 10$**. This is because the BHs lose the memory of their initial assembly after they grow well above this initial seed mass. In particular, the heavy seed hosts become incorporated into massive, metal-enriched galaxies, similar to those hosting the SMBHs that had grown from light seeds. The special conditions for the BH growth which existed in the ACH are destroyed, and rapid star formation is enabled. The subsequent stochastic growth of the BH and its host galaxy quickly brings the BH-host relations in line with those expected for light-seeded BHs which have been stochastically evolving for a long time.

X-ray emission is the best way to detect high-redshift SMBH seeds, for several reasons: luminous X-ray emission is ubiquitous from actively accreting BHs; it is relatively insensitive to obscuration effects, especially at high rest-frame energies; and there is a large contrast between BH accretion light and starlight in the X-ray band, so the two are not easily confused.

*Proposed Experiment.* Four of the six observational diagnostics of the BH seed models discussed above can be implemented via a combination of a sensitive X-ray survey which identifies high-$z$ BHs accreting at ≈ Eddington rate, and a sensitive OIR survey which characterizes their host galaxies:

- The sensitivity in the X-rays should reach down to $M_{BH} \approx 10^4\, M_\odot$ BHs at $z = 10$, or $f_x \approx 10^{-19}$ erg s$^{-1}$ cm$^{-2}$. At this sensitivity, a significant number of detections is expected in all models [55, 65], and redshifts $z \gtrsim 10$ where large differences in the luminosity function are generally expected [55]. At somewhat lower redshifts ($z \approx 8 - 9$), such a survey provides good sampling of the faint end of the luminosity function, which also is a constraint on seed models [56, 57]. Source positions should be known to a fraction of 1″ to enable unique IDs with potential host galaxies.

- To probe the expected differences in the host galaxy properties, the sensitivity in the OIR should be sufficient to identify counterparts of detected X-ray sources at least down to the mass limits expected in the light seed case. In the heavy seed case, where neither the host galaxies, nor the recently born ~ $10^5\, M_\odot$ seed BHs are expected to be detectable even with JWST, it will be possible to isolate the corresponding population as (brighter) X-ray point sources with no OIR counterparts (see Figure on next page). Furthermore, it will be possible to test the offsets between BHs and their bright galactic neighbors, anticipated in the close vicinity (up to tens of kpc or a few arcsec) of faint heavy seed hosts [36, 41, 79].

- There is a reasonable chance of catching on-going direct collapse events in such a survey (this possibility is considered separately in [74]). In this case, spectral signatures may be detectable (including a strong Ly$\alpha$ line [72]). Finally, a sensitive survey we describe is an excellent dataset for cross-correlations of the residual X-ray background with future 21-cm surveys [67].

*Capability Requirements.* The basic requirements for such a program include the following:

- *A sensitivity sufficient to detect $10^4\, M_\odot$ BHs at $z = 10$.* For reasonable assumptions about underlying X-ray spectral shapes, expected levels of obscuration, and bolometric corrections, the

# ELECTROMAGNETIC WINDOW INTO THE DAWN OF BLACK HOLES

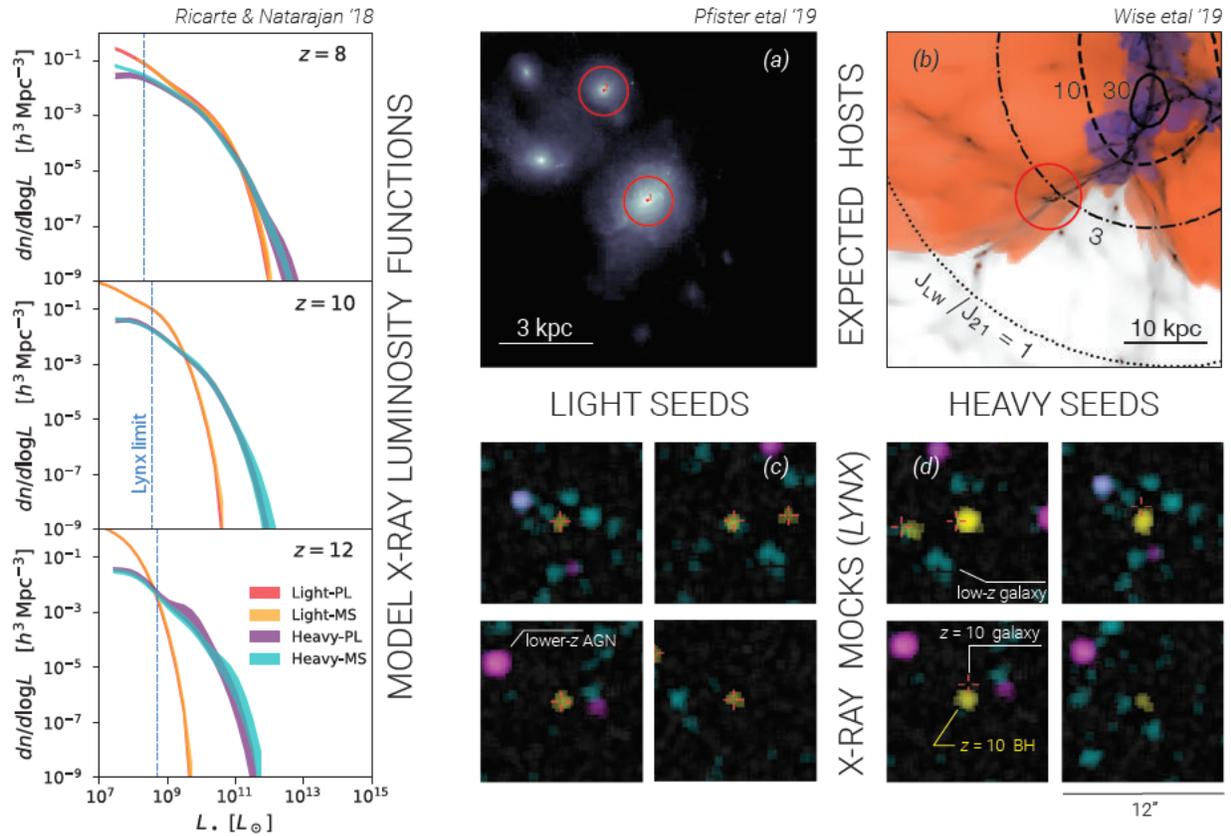

Future OIR and X-ray surveys will be able to reach into the $z = 10$ Universe. This is sufficiently close to the formation epoch of the first massive black holes. Therefore, we expect distinct observable differences between light and heavy-seeded models in the distribution of BH masses (and hence, luminosities) and in the properties of their host galaxies. Left: model X-ray luminosity functions (adopted from [55], RN18b hereafter). There are large differences at $z \geq 10$ for high $L_X$, which disappear by $z = 8$ when BHs "lose" memory of their initial assembly. Top panels: Locations of the BHs in their immediate galactic environment when they reach $M_{BH} \sim 10^5 \, M_\odot$ (red circles). (a) Light-seeded BHs are within massive, $M_{tot} \approx 10^{10} \, M_\odot$, galaxies detectable in the IR with JWST or WFIRST. (b) Newly born heavy-seeded BHs are in low-mass atomic-cooling halos located near (tens of kpc) strong sources of Lyman-Werner radiation such as groups of galaxies [41]. The initial host is udetectable in the IR, except perhaps by stacking on a large number of X-ray locations. Lower panels: 12" cutouts around four brightest $z = 10$ BHs in a 3' × 3' region in a 4Msec Lynx exposure. The X-ray mocks start with the Illustris TNG light cone simulation [78], which provides NIR magnitudes, star formation rates (= $L_X$ from X-ray binaries [80]), and SMBH accretion rates for $z < 8$ galaxies. BHs at $z = 10$ are added using the RN18b model. (c) In the light-seeds case, X-ray detections (yellow) are co-located with OIR-detectable galaxies (locations marked by red crosses). (d) In the heavy-seeds case, we expect fewer but brighter X-ray detections. By the time we catch these BHs in a survey, they typically have grown, but are still within small, $M_{tot} \approx 10^8 \, M_\odot$ and faint, $M_\star \approx 4 \times 10^5 \, M_\odot$, galaxies which are often about to merge with a larger galaxy 5–10 kpc away [73]. The host remains undetectable (lower-right panel), but the larger nearby galaxy can be seen as an OIR counterpart noticeably offset ($\sim 1''$) from the X-ray position (other three panels). The Galactic or low-redshift origin of the X-ray source in this case can be ruled out by X-ray spectral hardness ratios and JWST magnitude limits.

required observed-frame 0.5–2 keV fluxes (corresponding to penetrating rest-frame $\approx$ 5–20 keV X-rays) are $\approx 10^{-19}$ erg cm$^{-2}$ s$^{-1}$. These fluxes are just above the observability floor set by the X-ray binary populations expected in high-$z$ galaxies [56]. We discussed above that in the heavy seed scenarios there is likely a population of X-ray sources without a co-located OIR counterpart. Their reliable identification requires extremely high sample purities which are typically achieved at flux levels ≈ a few × the nominal detection threshold. A survey threshold requirement of $10^4\,M_\odot$ ensures that the necessary purity is achieved for $10^5\,M_\odot$ BHs expected in the heavy seed scenarios.

- *Sufficient angular resolution to protect against X-ray source confusion by the large numbers of foreground galaxies.* Current estimates [e.g. 56, 80] forecast a sky density for such sources of $(120\text{–}450)\times 10^3\,\text{deg}^{-2}$ at the relevant X-ray flux levels. Avoiding source confusion in this regime requires a $\lesssim 1''$ (50% power diameter) point spread function. Having a coarser PSF quickly destroys the ability to detect X-ray sources at the required flux levels. Sub-arcsec angular resolution is also essential for accurate source positions required for matching X-ray detections with their host galaxies.

- *Sufficiently large solid angle to conservatively probe an expected range of BH occupation fractions.* At $z \approx 10$, the space density of potential heavy seed hosts ($M_\text{halo}$ =few$\times 10^7 M_\odot$) is $\approx 1\,\text{Mpc}^{-3}$. More massive hosts of light-seeded $10^{4-5}\,M_\odot$ BHs ($M_\text{halo} \sim 10^{10}\,M_\odot$) have a number density of $\approx 10^{-3}\,\text{Mpc}^{-3}$. To provide a sufficient cushion to model uncertainties, we suggest a 1 deg$^2$ survey as a reasonable target. Such a survey covers $6.5\times 10^6\,\text{Mpc}^3$ per $\Delta z = 1$ around $z = 10$. It will be sensitive for occupation factors of actively accreting BHs down to $f = f_\text{occ} \times f_\text{duty} \approx 10^{-6}$ and $10^{-3}$ for heavy- and light-seed models, respectively (corresponding to $\sim$ 10 sources per deg$^2$). These factors are 2 orders of magnitude below the upper end of the range predicted in several studies (e.g., [41, 44, 55, 66, 79, 81]), $f_\text{heavy} \approx 10^{-4}$ and $f_\text{light} \approx 10^{-1}$, which would result in $\sim$ 1000 sources per deg$^2$. Ideally, the survey should be conducted across 3–5 distinct fields that are widely separated in the sky to minimize, and allow assessment of, the effects of cosmic variance [e.g. 82, 83]. Some optimizations will be possible with a "wedding cake" survey strategy, but the exact design is mission-specific and thus not considered here.

- *Sensitivity in the OIR to characterize host galaxies, or lack thereof.* As discussed above, the OIR survey should reach sensitivities sufficient for detecting at least the light-seed hosts at $z = 10$, and reliably determining their photo-$z$. These hosts galaxies are expected to be faint, with near-infrared magnitudes of $\approx 28.5 - 29.5$. Such magnitudes should be reachable over the required solid angles by *JWST* "wide" and *WFIRST* "deep" surveys in well-studied multiwavelength survey regions [e.g. 84]. Maintaining a low fraction (< 1%) of low-redshift interlopers is essential, as is the case with studies of high-$z$ galaxy populations in general. Photometric-redshift and Lyman-break techniques have promise for enabling such a challenging discrimination, but ultradeep multi-band OIR imaging and optimal redshift-estimation approaches will be essential.

The OIR requirements outlined above can be met with the already planned *JWST* and *WFIRST* missions (although adustments to the currently discussed observing plans may be needed). The X-ray requirements, however, can be met only with a new, next-generation X-ray observatory. To conduct the required deep survey in less than $\sim$ 1 year of total exposure time, such an observatory should combine high throughput (at least 1–2 m$^2$ effective area in the soft X-ray band) with sub-arcsec angular resolution maintained over a large field-of-view. *Chandra* X-ray Observatory's grasp falls short of the requirements by $\sim$ three orders of magnitude. *Athena*'s sensitivity is insufficient by a factor of 200 due to its 5$''$ PSF. The proposed *Lynx* mission concept uniquely meets all of the requirements.

***Conclusions.*** In the electromagnetic bands, the best constraints on the origin of the earliest SMBH can come from directly detecting their accreting seeds in X-ray observations. The required observations are obtainable with a next-generation, high angular resolution X-ray telescope. Combined with OIR observations of their host galaxies and their immediate galactic environment, they have a great potential to solve the long-standing puzzle of the origin of SMBHs. The experiment suggested here is highly synergistic with the future gravitational wave studies: while the X-ray signal probes accretion (but is blind to mergers), *LISA* will be able to directly track mergers (but will be blind to accretion). These two probes together can unveil a complete picture of the early SMBH assembly.